\documentclass[pre,aps,twocolumn,floatfix,superscriptaddress]{revtex4}
\usepackage{amsmath}
\usepackage{color}
\usepackage{bbm}
\usepackage{graphicx}

\begin{document}

\title{OpenStreetCab: Exploiting Taxi Mobility Patterns in New York City to Reduce Commuter Costs}
\author{Vsevolod Salnikov}
\affiliation{naXys, University of Namur, Belgium}
\author{Renaud Lambiotte}
\affiliation{naXys, University of Namur, Belgium}
\author{Anastasios Noulas}
\affiliation{ComputerLab, University of Cambridge, UK}
\author{Cecilia Mascolo}
\affiliation{ComputerLab, University of Cambridge, UK}

\maketitle

The rise of Uber as the global alternative taxi operator has attracted a lot of interest
recently. Aside from the media headlines which discuss the new phenomenon, e.g. on how it has disrupted
the traditional transportation industry, policy makers, economists, citizens and scientists  
have engaged in a discussion that is centred around the means to integrate the 
new generation of the sharing economy services in urban ecosystems. In this work, 
we aim to shed new light on the discussion,
by taking advantage of a publicly available longitudinal dataset that describes the mobility
of yellow taxis in New York City. In addition to movement, this data contains information 
on the fares paid by the taxi customers for each trip. As a result we are given the 
opportunity to provide a first head to head comparison between the iconic yellow taxi 
and its modern competitor, Uber, in one of the world's largest metropolitan centres. 
We identify situations when Uber X, the cheapest version of the Uber taxi service, tends to be more 
expensive than yellow taxis for the same journey. We also demonstrate how Uber's
economic model effectively takes advantage of well known patterns in human movement. Finally, we take
our analysis a step further by proposing a new mobile application that compares taxi prices in the city
to facilitate traveller's taxi choices, hoping to ultimately to lead to a reduction of commuter costs. 
Our study provides a case on how big datasets that become public
can improve urban services for consumers by offering the opportunity for transparency in economic
sectors that lack up to date regulations.

\section{Taxi Price Comparison Experiment}

\begin{figure}
  \centering
    \includegraphics[width=0.7\linewidth]{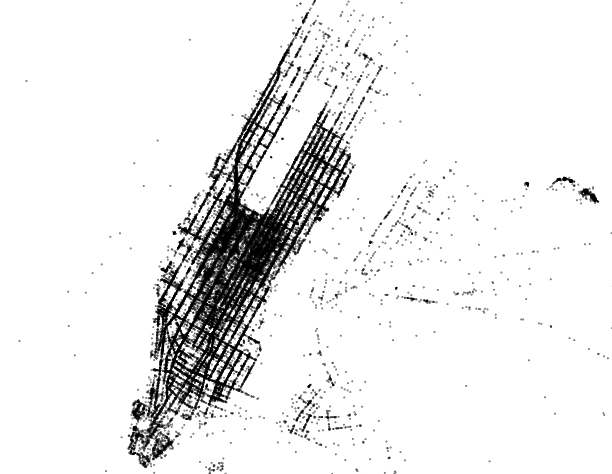}
\caption{Marking the traces of new york city yellow taxis. For every pick up and drop off point in a uniform sample of the data we draw a black point.}  
\label{taxitraces}  
\end{figure}

\paragraph*{\textbf{The New York City Taxi Dataset.}}
The Freedom of Information Law in United States encourages public authorities 
to release their data where appropriate to the benefit of the citizens. In 2014 
the law was exploited by Chris Whong to acquire and post on the web one of the most 
comprehensive taxi mobility datasets available today. The dataset describes
taxi journeys in New York City during the full course of 2013, and informs us not only on the
origin and destination points of taxi trips, noted in the related jargon as pick up and 
drop off points respectively, but also on the financial costs incurred to the customer (trip fair)
with unprecedented detail. This rather dense mobility dataset, containing hundreds of millions
of trips is of gigabytes in size and can be downloaded here~\url{http://chriswhong.com/open-data/foil_nyc_taxi/}. A sample of the traces generated
by the data can be seen is in Figure~\ref{taxitraces}, where we have drawn a black point for every pick up
and drop off point of a taxi journey. 

\paragraph*{\textbf{Comparing Taxi Prices}}
In August 2014, Uber opened up an API with access to valuable information about its services.
The occasion allowed us to perform a first head to head comparative analysis of prices between Uber and Yellow taxis 
in New York City. To achieve this we run the following experiment : 
\begin{itemize}
\item 1. For every trip in the New York City Yellow Taxi dataset, record the geographic coordinates (latitude and longitude) of the pick up and drop off points. 
\item 2. Retrieve the total fare paid by the customer for the trip (including the tip).
\item 3. Query Uber's API and ask how much they would charge for the same trip (same pick up and drop off points), 
considering the cheapest version of the service, Uber X.
\item 4. Uber's API returns a value range indicating the minimum and maximum price estimate. We take the mean
of the two values.
\item 5. We then compare the prices from the two services. 
\end{itemize}

\begin{figure}
  \centering
    \includegraphics[width=0.7\linewidth]{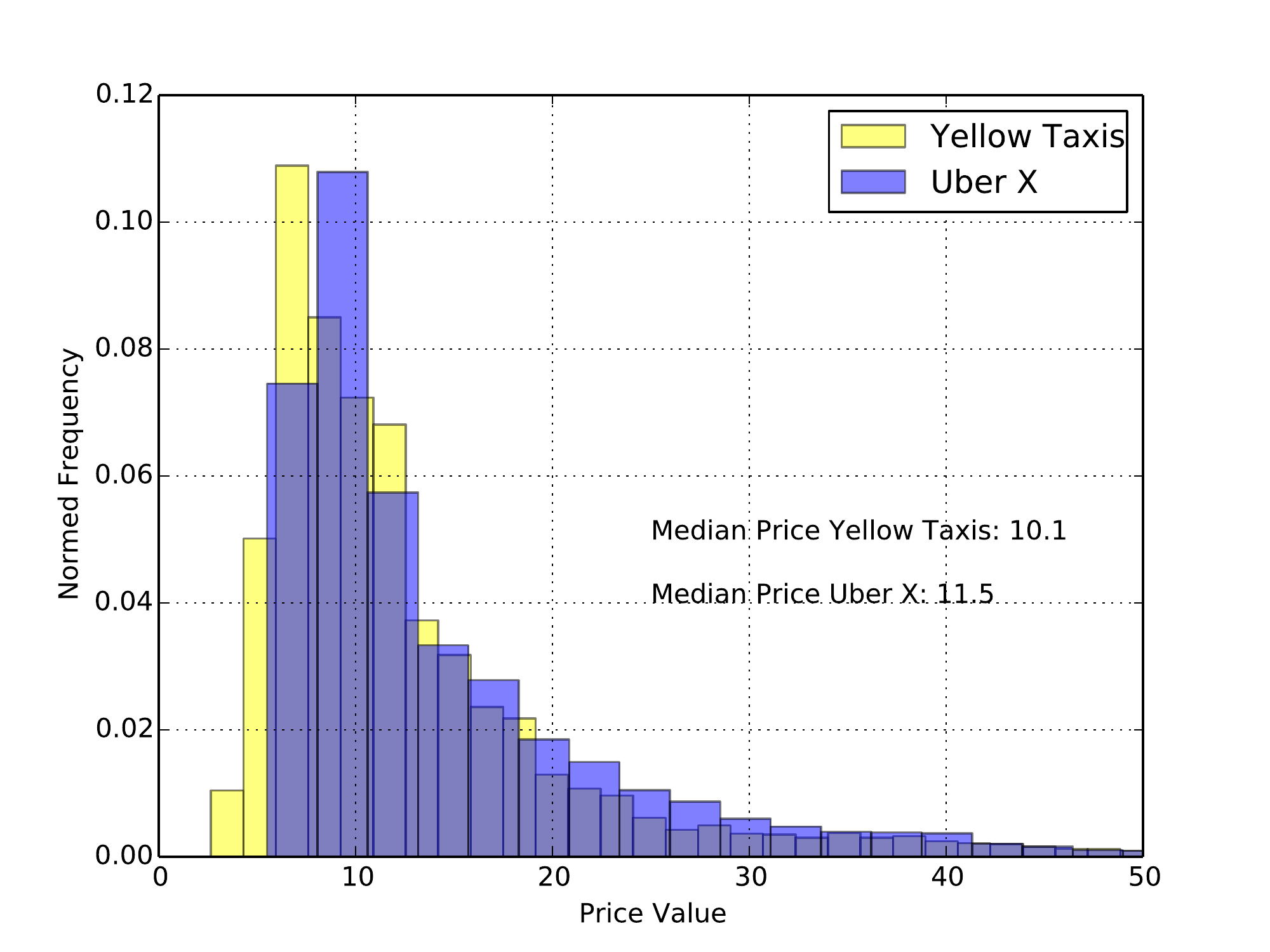}
\caption{Distribution of prices per journey for Uber X and Yellow Taxis in New York City.}  
\label{pricecomp}  
\end{figure}

\begin{figure}
  \centering
    \includegraphics[width=0.7\linewidth]{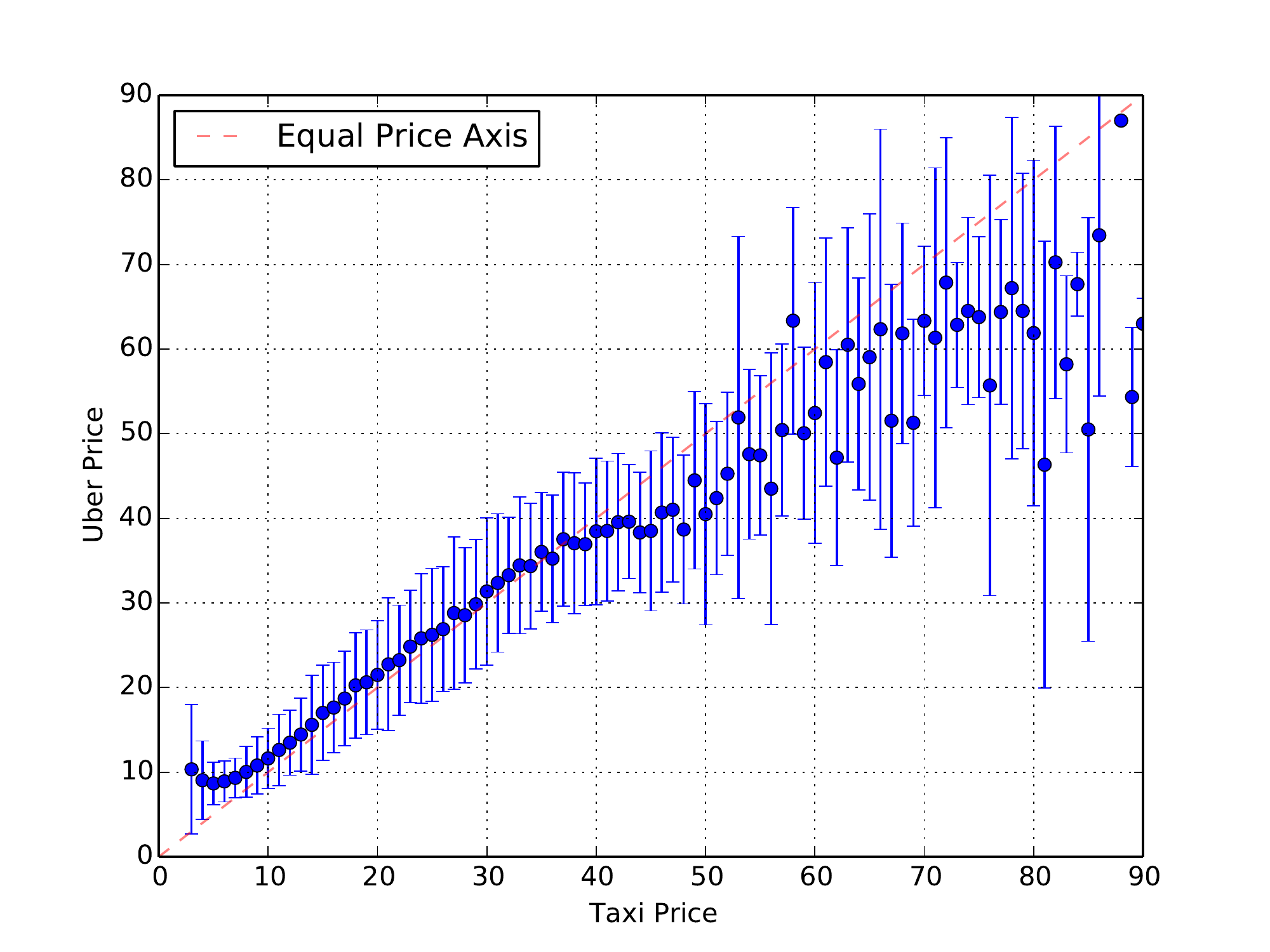}
\caption{Median Uber price for a given Yellow Taxi price.}  
\label{ubervsyellow}  
\end{figure}

\begin{figure}
  \centering
    \includegraphics[width=0.7\linewidth]{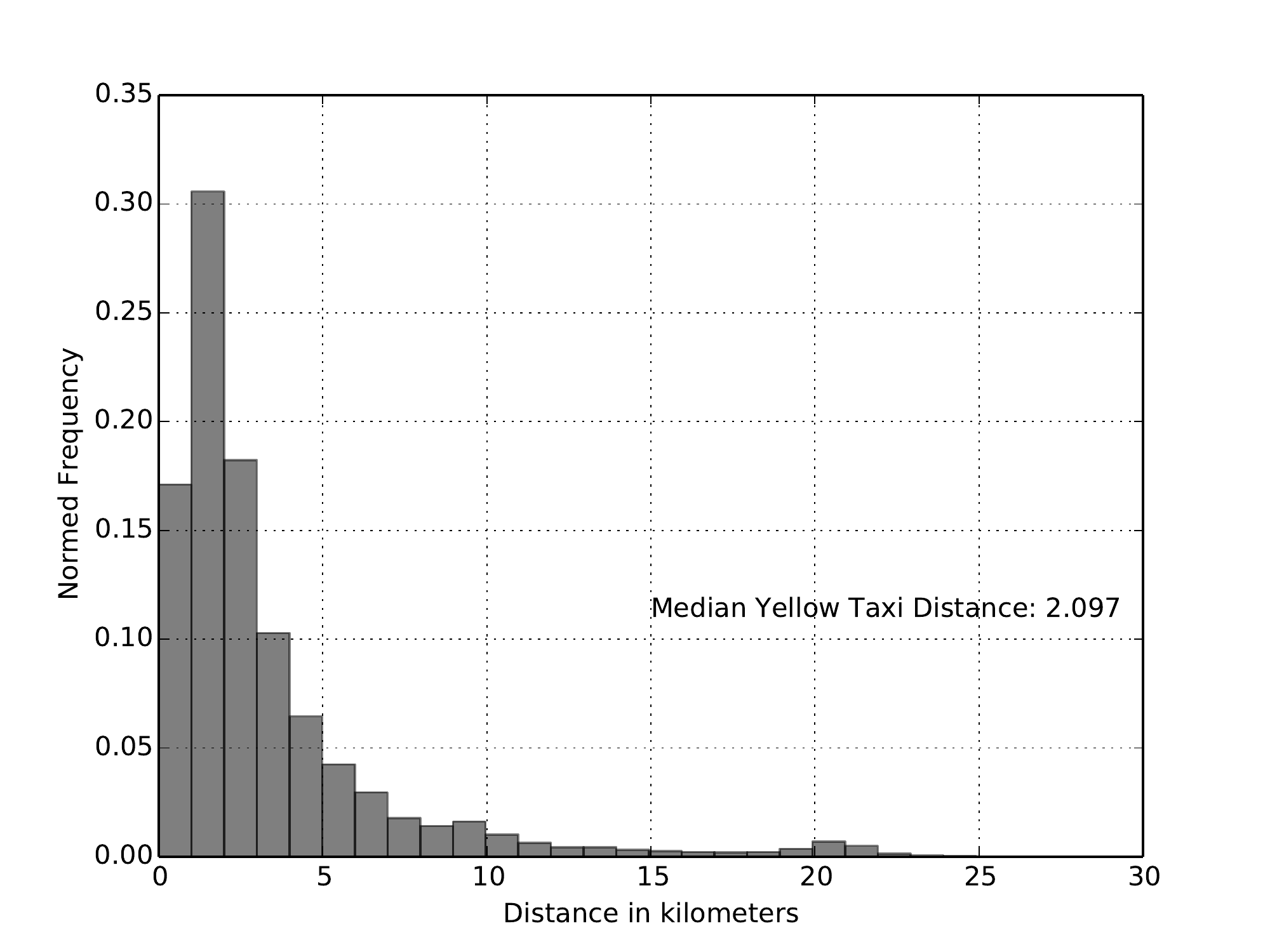}
\caption{Distribution of geographic distances between drop off and pick up points for Yellow Taxi journeys.}  
\label{tripdistances}  
\end{figure}

As can be observed in Figure~\ref{pricecomp} where the distribution of prices for the two services is shown, despite the qualitative similarity of the two distribution, yellow taxi appear on average (median) 1.4 US dollars cheaper than Uber X. In Figure~\ref{ubervsyellow}, we compare Uber and yellow taxis from another perspective: for every observed yellow taxi price, we show the median Uber X price. Uber appears more expensive for prices below 35 dollars and begins to become cheaper only after that threshold. As one would expect, the cheaper journeys are those that are in principle of shorter range. As observed in a variety of empirical data, human mobility tends to be characterised by a vast majority of short trips~\cite{gonzalez2008understanding, brockmann2006scaling}. This observation therefore suggests that Uber's economical model exploits this trend of human mobility in order to maximise revenue. We also confirm the skewed frequency distribution of movement distances in the present context by visualising it in Figure~\ref{tripdistances}, where we note a mean distance for a yellow taxi trip in New York equal to $2.09$. 

The above experiment may involve a number of biases which we refer to here. 
The NYC Yellow taxi data corresponded to year 2013 whereas Uber to 2014. Although note that the prices for yellow taxis in the city had last changed in 2012 after 8 years~\cite{fares}. So it should offer a good approximation of today’s prices. Further, there was no control for time of the day/week for the API query, an additional dimension which should be incorporated when available. However, we argue that the process of comparing two different companies that provide the same service in the same geographic area is of value to commuters. Just as consumer have open access to airfares for a long time now allowing for transparency in a free, competitive, market we believe that similar approaches could benefit commuters in modern cities.  

\begin{figure}
  \centering
    \includegraphics[width=0.7\linewidth]{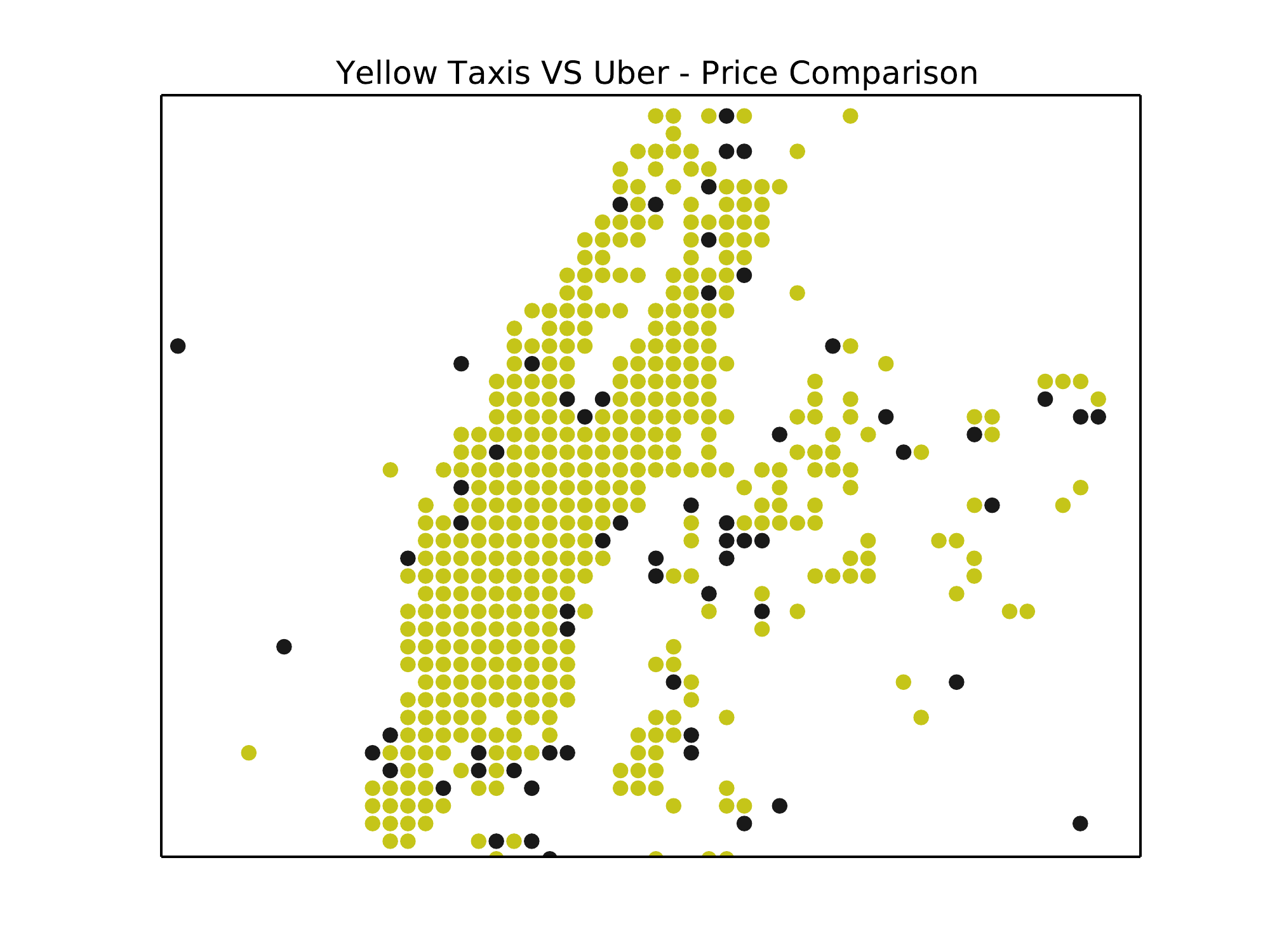}
\caption{Geographic comparison between Uber and Yellow Taxi prices. We paint an area black if Uber is cheaper by trip majority and yellow otherwise.}  
\label{ios1}  
\end{figure} 

\section {Helping commuters}

Our observations show that it might be financially advantageous on average for travellers to chose either Yellow Cabs or Uber depending on the duration of their journey. However the specific journey they are willing to take matters.
In order to help users to take the right decision, we have developed a smartphone app, called OpenStreetCab, designed as follows.

One limitation for the design of our service is that only prices for trips with origins and destinations 
in the New York City Taxi Dataset can in principle be retrieved. In order to evaluate the price of any trip, as needed for a usable App, we have divided the NY region into a mesh with cells of size around 100m by 100m in order to index trips in the database efficiently. For each user query, we find a set of trips in our dataset with the origin in neighbouring cells of desired origin and, among them, we find the trip whose destination is closest to the desired one. This strategy has the advantage of being sufficiently fast to perform online queries and expected to provide reliable price estimates. For the same trip, Uber price is obtained through their API.

\begin{figure}
  \centering
    \includegraphics[width=0.7\linewidth]{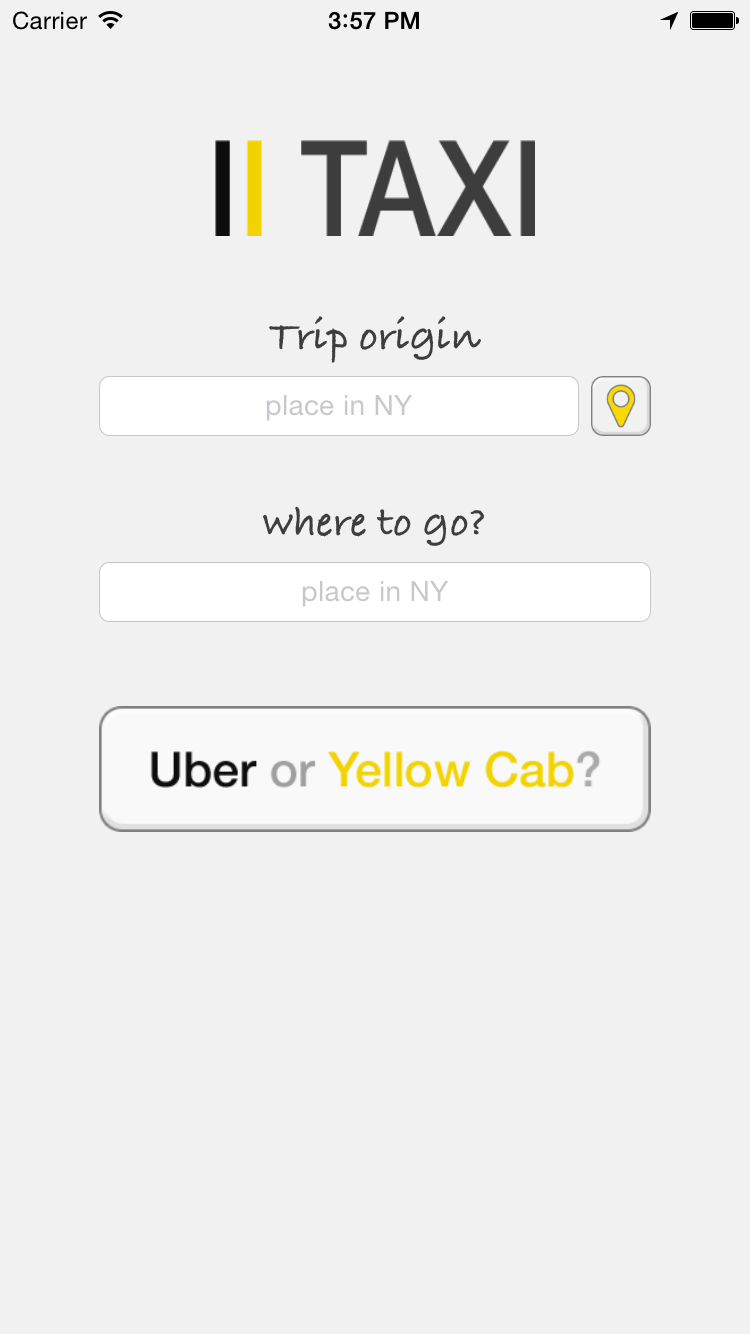}
\caption{The proof of the concept}  
\label{ios1}  
\end{figure} 

A real-time prototype has been designed and is currently launched on popular mobile platforms. Future improvements include the possibility to change predictions depending on the time of the day, or on the expected traffic on the way, but also to suggest other types of transportations, such as walking when the distance is sufficiently short, or only part of the way, in situations when a small change in the origin point can lead to a significant change in the price quote. In the meanwhile the current version (Fig. \ \ref{ios1}) already provides a fully working solution, including geolocation services and address retrieval. We are planning to launch the application on the related stores very soon.

\small
\bibliographystyle{plain}
\bibliography{biblio}

\begin{thebibliography}{1}

\bibitem{brockmann2006scaling}
Dirk Brockmann, Lars Hufnagel, and Theo Geisel.
\newblock The scaling laws of human travel.
\newblock {\em Nature}, 439(7075):462--465, 2006.

\bibitem{gonzalez2008understanding}
Marta~C Gonzalez, Cesar~A Hidalgo, and Albert-Laszlo Barabasi.
\newblock Understanding individual human mobility patterns.
\newblock {\em Nature}, 453(7196):779--782, 2008.

\bibitem{fares}
Taxi Fares~To Rise.
\newblock {New York Times}, 2012.
\newblock
  \url{http://cityroom.blogs.nytimes.com/2012/07/12/taxi-fares-in-new-york-to-rise-by-17/?_r=0}.

\end{thebibliography}
\end{document}